\pdfoutput=1
\documentclass[aps,pra,superscriptaddress,twocolumn]{revtex4-2}
\usepackage[utf8]{inputenc}
\usepackage{amsmath,amssymb,amsthm,mathrsfs,bm}
\usepackage{lmodern}
\usepackage{graphicx}
\usepackage{braket}
\PassOptionsToPackage{hyphens}{url}
\usepackage{units,hyperref}
\usepackage{dsfont}
\usepackage{qcircuit}
\usepackage{booktabs}
\usepackage{orcidlink}

\theoremstyle{plain}
\newtheorem{theorem}{Theorem}
\newtheorem{remark}[theorem]{Remark}
\newtheorem{innerrestatethm}{Theorem}
\newenvironment{restatethm}[1]
  {\renewcommand\theinnerrestatethm{#1}\innerrestatethm}
  {\endinnerrestatethm}

\numberwithin{equation}{section}

\begin{document}

\title{Experimental simulation of a Hilbert-space inner-product change}

\author{Vinh San Dinh\orcidlink{0000-0001-9258-9542}}
\affiliation{Superconducting Quantum Materials and Systems Center (SQMS), Fermi National Accelerator Laboratory, Batavia, IL 60510, USA}
\affiliation{Northwestern-Fermilab Center for Applied Physics and Superconducting Technologies,
Northwestern University, Evanston, Illinois 60208, USA}
\affiliation{Graduate Program in Applied Physics, Northwestern University, Evanston, Illinois 60208, USA}

\author{Salini Karuvade\orcidlink{0000-0002-1513-7857}}
\affiliation{Institute for Quantum Science and Technology, University of Calgary, Alberta, Canada T2N 1N4}
\affiliation{Department of Physics, Concordia University, Montreal, QC H4B 1R6, Canada}

\author{Abhijeet Alase\orcidlink{0000-0002-5230-4597}}
\affiliation{Institute for Quantum Science and Technology, University of Calgary, Alberta, Canada T2N 1N4}
\affiliation{Department of Physics, Concordia University, Montreal, QC H4B 1R6, Canada}

\author{Silvia Zorzetti\orcidlink{0000-0002-3208-3387}}
\affiliation{Superconducting Quantum Materials and Systems Center (SQMS), Fermi National Accelerator Laboratory, Batavia, IL 60510, USA}

\author{Barry C. Sanders\orcidlink{0000-0002-8326-8912}}
\email{sandersb@ucalgary.ca}
\affiliation{Institute for Quantum Science and Technology, University of Calgary, Alberta, Canada T2N 1N4}
\date{November 15, 2026}
\begin{abstract}
We simulate an effective change in a single qubit's inner product (IP) by implementing the channel $\mathcal{F}^\eta$ on a superconducting quantum processor. Whereas previous experiments simulated parity-time (PT)-symmetric dynamics with a fixed IP, this experiment implements the IP-change channel itself for metrics associated with PT-symmetric systems. Building on an earlier adversarial verification approach, we derive a process-fidelity threshold that rules out mixtures of qubit unitary operations with random discarding of runs. We test the channel for 40 metrics using a single qutrit and a two-qubit Stinespring dilation. The measured fidelities exceed the threshold in 24 of 40 qutrit instances after readout error mitigation, compared with 16 of 40 unmitigated two-qubit instances. Without mitigation, 14 of 40 qutrit instances pass. These results demonstrate a direct simulation of an effective Hilbert-space IP change and its verification against a specified adversarial class, while showing the importance of readout treatment in comparing the encodings.
\end{abstract}
\maketitle 

\section{Introduction}
\label{sec:intro}
The inner product (IP) defines the geometry of Hilbert space and determines the probabilities predicted by quantum mechanics~\cite{SN21}. Although the IP is usually treated as fixed, certain non-Hermitian Hamiltonians in parity-time (PT)-symmetric quantum mechanics generate unitary evolution with respect to a modified IP~\cite{BB98, BBJ02, Mos02}. Recent theoretical work describes IP changes as trace-non-increasing quantum channels within standard quantum mechanics~\cite{BBJM07,BBC+13,KAS22,AKS22,Kar22}. These channels provide an operational route to simulating IP changes, which we test experimentally here.

PT-symmetric quantum dynamics have been simulated in photonic, atomic and superconducting systems~\cite{BH24, EGM+18, ORN+19, BBK21, AGU20, NAJ+19, CAJ+21, CAH+22a, LHL+19, DSZ+21, WZZ+21, RLZ+22, XZB+17, MYW+22, HWH+23}.
These studies simulate evolution generated by effective non-Hermitian Hamiltonians while keeping the underlying IP fixed. In contrast, our focus is on implementing the IP-change channel itself, rather than simulating PT-symmetric dynamics.

We implement the single-qubit IP-change channel $\mathcal{F}^\eta$ on Rigetti's Aspen-11 superconducting processor using two encodings: a single qutrit and a two-qubit Stinespring dilation. Both realize the channel through coherent coupling to auxiliary degrees of freedom followed by post-selection. The qutrit uses its third energy level, while the two-qubit encoding uses an ancilla and entangling gates. Because random discarding can also reproduce probability loss, observing loss alone does not establish that either implementation realizes the target channel. Verification must therefore distinguish the target operation from mixtures of qubit unitary operations with random discarding of runs.

Building on the adversarial verification approach of Ref.~\cite{KAS22}, we derive a process-fidelity form of its threshold and apply it to both experimental encodings. The bound gives the highest fidelity achievable by the specified adversarial class, so a true process fidelity above it rules out those strategies. For each sampled metric, we compare the fidelity estimated by process tomography with this threshold. The qutrit encoding passes more instances after readout error mitigation, which was applied only to its data; without mitigation, the two-qubit encoding passes slightly more. This distinction motivates a comparison of both raw and mitigated results below.

\section{Mathematical Framework}
\label{sec:mathematicalframework}
A Hilbert-space IP change can be represented by a trace-non-increasing quantum channel~\cite{KAS22}. The following subsections describe its implementation using a single qutrit or two qubits. The benchmark bounds the fidelity achievable by mixtures of qubit unitary operations with random discarding of runs. A process fidelity above this bound rules out those strategies.

\subsection{Concept: IP Changes as Physical Processes}
\label{subsec:opframework}
IP changes can be implemented as physical operations and arise in descriptions of PT-symmetric systems~\cite{KAS22}.

Let~$\mathcal{H}$ be a finite-dimensional Hilbert space with the standard inner product~$\braket{\bullet|\bullet}$. For any unbroken PT-symmetric Hamiltonian $H_\text{PT}$, there exists a positive-definite metric operator,
\begin{equation}
\eta\in\mathcal{B}(\mathcal{H}),
\end{equation}
such that
\begin{equation}
\eta^{-1}H_\text{PT}^\dagger\eta = H_\text{PT}.
\end{equation}
This quasi-Hermiticity condition~\cite{Mos03a} implies that $H_\text{PT}$ is Hermitian with respect to the modified IP
\begin{equation}
\braket{\bullet|\bullet}^\eta := \braket{\bullet|\eta|\bullet},
\end{equation}
where $\eta$ is not unique.

The IP change is represented by a completely positive, trace-non-increasing map $\mathcal{F}^\eta$. The condition $\eta\leq\mathds{1}$ ensures that the operation does not increase the trace. Its action is
\begin{equation}
\mathcal{F}^\eta(\rho) = F^\eta \rho F^{\eta\dagger}, \quad \text{where} \quad F^\eta := \eta^{1/2}.
\end{equation}

Acting on pure states $\ket{\psi}, \ket{\phi} \in \mathcal{H}$, this map induces
\begin{equation}
\ket{\psi} \mapsto \eta^{1/2}\ket{\psi}, \quad \ket{\phi} \mapsto \eta^{1/2}\ket{\phi},
\end{equation}
such that their IP transforms as
\begin{equation}
(\bra{\phi}\eta^{1/2})(\eta^{1/2}\ket{\psi}) = \braket{\phi|\eta|\psi} = \braket{\phi|\psi}^\eta.
\end{equation}

IP transformations also provide a way to simulate unbroken PT-symmetric dynamics. For a non-Hermitian Hamiltonian $H_\text{PT}$ with unbroken PT symmetry, define $\tilde{U}_\text{PT}:=\text{e}^{-\text{i}H_\text{PT}t/\hbar}$. This evolution is unitary with respect to the modified IP, but generally not the standard IP. The scaled evolution $\sqrt{\kappa}\tilde{U}_\text{PT}$ consists of an initial IP transformation, evolution under a Hermitian Hamiltonian, and a scaled inverse IP transformation:
\begin{equation}
\sqrt{\kappa}\tilde{U}_\text{PT} =F^{\kappa\eta^{-1}} \text{e}^{-\text{i} h_\text{PT}t/\hbar}F^{\eta},
\end{equation}
where $h_\text{PT}=\eta^{1/2}H_\text{PT}\eta^{-1/2}$ is Hermitian. The largest allowed scaling factor is $\kappa=1/\|\eta^{-1}\|=r$, where $r$ is the minimum eigenvalue of $\eta$. This choice keeps the operation defined by $F^{\kappa\eta^{-1}}$ trace-non-increasing~\cite{KAS22}. The present experiment implements and benchmarks $\mathcal{F}^\eta$.

\subsection{Construction: Channel and Dilation}
\label{subsec:channel_dilation}

Unitary dilation and post-selection implement the IP-change channel. For a qubit state $\rho\in\mathcal{B}(\mathcal{H}_2)$ and metric $\eta\in\mathcal{B}(\mathcal{H}_2)$, the channel acts as
\begin{equation}
\mathcal{F}^\eta(\rho) = F^\eta \rho F^{\eta\dagger},
\end{equation}
where $F^\eta=\eta^{1/2}$ is its Kraus operator. The condition $\eta\leq\mathds{1}$ gives $F^{\eta\dagger}F^\eta\leq\mathds{1}$, so the channel is trace-non-increasing.

The dilation embeds the input as $\tilde{\rho}$ in a larger Hilbert space. A unitary $U^\eta$ acts on this space, including the auxiliary degrees of freedom. Projection onto the original qubit subspace must recover the target Kraus operator. With $P$ denoting this projector, the requirement is
\begin{equation}
\label{eq:unitary_projection}
P U^\eta P = \eta^{1/2}.
\end{equation}
Applying $U^\eta$ and retaining the outcome selected by $P$ gives
\begin{equation}
\label{eq:channel_action_dilated}
\mathcal{F}^\eta(\rho) = P \left( U^\eta \tilde{\rho} U^{\eta\dagger} \right) P.
\end{equation}

\subsection{Implementation: Encodings}
\label{subsec:implementation}
Both encodings implement $\mathcal{F}^\eta$ for the same input state $\rho\in\mathcal{B}(\mathcal{H}_2)$ and metric $\eta$. The metric has eigenvalues $1$ and $r$ and angular coordinates $(\theta,\phi)$. The encodings differ in their auxiliary degrees of freedom.

\subsubsection{Single-Qutrit Implementation}

The qutrit encoding uses $\ket{0}$ and $\ket{1}$ for the target qubit and $\ket{2}$ as the auxiliary level. It requires no entangling gates. The embedded input is $\tilde{\rho}_3=\rho\oplus0\in\mathcal{B}(\mathcal{H}_3)$.

The unitary $U_3^\eta$ satisfies $PU_3^\eta P=\eta^{1/2}$ with $P=\mathds{1}_2\oplus0$. Its block form is
\begin{equation}
\label{eq:U3eta}
U^\eta_3 = \begin{pmatrix} 
\eta^{1/2} & \bm{u} \\ 
-\text{e}^{\text{i} \xi}\bar{\bm{u}}^{\top} & \text{e}^{\text{i}\xi}\sqrt{r} 
\end{pmatrix},
\end{equation}
where $\sqrt{r}$ is the smallest eigenvalue of $\eta^{1/2}$. The vector $\bm{u}\in\mathbb{C}^2$ lies along the corresponding eigendirection and has norm $\|\bm{u}\|=\sqrt{1-r}$. Writing $\bm{u}=\text{e}^{\text{i}\varphi}\sqrt{1-r}\,\bm{v}_r$ for a normalized eigenvector $\bm{v}_r$ makes its phase freedom explicit. Here, $\varphi$ is a dilation phase, distinct from the metric azimuth $\phi$. Appendix~\ref{app:qutrit_compilation} shows that the following gate sequence realizes this matrix with $\xi=\varphi=0$:
\begin{widetext}
\begin{equation}
\label{eq:3circuit}
\Qcircuit @C=1em @R=.7em {
& \gate{U^{\eta}_{3}} & \qw
}
=  
\begin{array}{c}
\Qcircuit @C=1em @R=.7em 
{
& \gate{R^{01}_{z}(-\phi)} & \qw & \gate{R^{01}_{y}(-\theta)} & \qw & \gate{R^{12}_y(-\gamma)} & \qw & \gate{R^{01}_y(\theta)} & \qw & \gate{R^{01}_z(\phi)} & \qw & \gate{P} & \qw &
}
\end{array}
\end{equation}
\end{widetext}
with $\gamma=2\cos^{-1}(\sqrt{r})$. The sequence transfers amplitude to $\ket{2}$, which projection excludes from the retained outcome. The ideal sequence introduces no residual phases in the qubit subspace (Appendix~\ref{app:qutrit_compilation}). Appendix~\ref{app:qutrit_compilation} also describes its compilation into microwave pulses and virtual $Z$-gates. This implementation requires pulse-level calibration of both transitions.

\subsubsection{Two-Qubit Stinespring Dilation}

The two-qubit encoding uses $Q_l$ for the target system and $Q_m$ as the ancilla. Its input state is $\tilde{\rho}_{2\times2}=\rho\otimes\ket{0}\bra{0}$.

Single-qubit rotations and native entangling gates implement $U_{2\times2}^\eta$, programmed in pyQuil~\cite{pyquil}. With $P=\mathds{1}_2\otimes\ket{0}\bra{0}$, the dilation has block form
\begin{equation}
U^\eta_{2\times2} = \begin{pmatrix} \eta^{1/2} & \text{i}(\mathds{1}_2-\eta)^{1/2} \\ \text{i}(\mathds{1}_2-\eta)^{1/2} & \eta^{1/2} \end{pmatrix}.
\end{equation}
Appendix~\ref{app:qubit_compilation} gives the native-gate compilation. The circuit is
\begin{widetext}
\begin{equation}
\label{eq:2circuit}
\begin{array}{c}
\Qcircuit @C=1em @R=.7em {
& \multigate{1}{U^{\eta}_{2\times2}} & \qw \\
& \ghost{U^{\eta}_{2\times2}} & \qw
}
\end{array}
=  \qquad 
\begin{array}{c}
\Qcircuit @C=1em @R=.7em 
{
& \ustick{Q_l}\qw &\gate{R_z(-\phi)} &\qw & \gate{R_y(-\theta)} & \qw & \targ & \multigate{1}{{\rm XY}(\gamma)} & \targ & \qw & \gate{R_y(\theta)} & \qw & \gate{R_z(\phi)} & \qw &
\\
\lstick{\ket{0}}  & \dstick{Q_m}\qw & \qw & \qw & \qw & \qw & \ctrl{-1} & \ghost{{\rm XY}(\gamma)} & \ctrl{-1} & \qw & \qw & \qw & \qw & \meter & \rstick{= 0}
}
\end{array}
\end{equation}
\end{widetext}
The parametric ${\rm XY}$ gate transfers amplitude to the ancilla subsystem. Its angle is $\gamma=2\cos^{-1}(\sqrt{r})$, as in the qutrit implementation (Appendix~\ref{app:qubit_compilation}). Post-selection on the ancilla outcome $\ket{0}$ gives the target channel.

Both encodings implement the same channel. The qutrit uses an auxiliary energy level; the two-qubit encoding uses an ancilla and entangling gates. The same adversarial benchmark tests both implementations.

\subsection{Benchmarking Framework and Dimension Witness}
\label{subsec:verification}

Probability loss alone does not establish that an implementation realizes the target channel. In a dilation, loss arises from post-selection on an auxiliary degree of freedom. Randomly discarding runs can also reduce the trace without implementing the intended process.

The adversarial benchmark draws on validation strategies in quantum key distribution and loophole-free tests of local realism~\cite{BLMS00,FS17}. The model allows mixtures of unitary operations within the qubit subspace and random discarding of runs. It excludes access to additional coherent degrees of freedom, such as higher energy levels or an entangled ancilla.

An allowed strategy has the form $\mathcal{F}^{\rm dis}_\eta(\rho)=\sum_{j=1}^M q_jU_j\rho U_j^\dagger$, where each $U_j$ is unitary and $q_j\geq0$ is the probability of applying $U_j$ and retaining the run. The remaining probability $1-\sum_jq_j$ corresponds to discarding the run. For the normalized process fidelity to be defined, the retained probability must satisfy $0<\sum_jq_j\leq1$. The following theorem bounds the process fidelity $\Delta(\cdot,\cdot)$ with the target channel $\mathcal{F}^\eta$.

\begin{theorem}\label{thm:thresh}
Let $\eta\le\mathds{1}_2\in\mathcal{B}(\mathcal{H}_2)$ be a metric operator with eigenvalues $\lambda_1 \ge \lambda_2$. Then, for any adversarial strategy $\mathcal{F}^{\rm dis}_\eta$ as defined above,
\begin{equation}
\Delta\left(\mathcal{F}^{\rm dis}_\eta,\mathcal{F}^\eta\right) \le D^{(2)}_\text{th}(\eta) := \frac{\sqrt{\lambda_1}+\sqrt{\lambda_2}}{\sqrt{2(\lambda_1+\lambda_2)}}.
\end{equation}
\end{theorem}
\noindent A full proof is given in Appendix~\ref{app:threshold}. For a metric with $\lambda_1 = 1$ and $\lambda_2 \equiv r \in (0,1]$, this reduces to the threshold used throughout this work,
\begin{equation}
\Delta_\text{th}^\eta = \frac{1+\sqrt{r}}{\sqrt{2(1+r)}}.
\end{equation}
The threshold depends only on the minimum eigenvalue $r$. As $r$ approaches $1$, the metric approaches the identity and the adversary can approximate the target more closely. The threshold therefore rises. At $r=1$, the threshold equals $1$, so no implementation can exceed it.

A process fidelity above the threshold rules out every unitary-mixture strategy in the defined adversarial class. Under these restrictions, the benchmark acts as a dimension witness: a test of the Hilbert-space dimension needed to implement a process~\cite{brunner2008testing}. Each sampled metric and encoding gives a separate test. An instance passes if its measured fidelity exceeds $\Delta_\text{th}^\eta$; otherwise, it fails. Experimental interpretation must account for uncertainty and systematic errors in the fidelity estimate, as discussed in Appendix~\ref{app:threshold}.

The benchmark also extends to IP-change channels in dimension $d$. Theorem~\ref{thm:thresh} and its proof in Appendix~\ref{app:threshold} treat the qubit case. The general threshold is $D^{(d)}_{\rm th}(\eta)=\operatorname{Tr}[\eta^{1/2}]/\sqrt{d\cdot\operatorname{Tr}[\eta]}$. The same rank-one Choi argument and von Neumann trace inequality apply in higher dimensions, with the corresponding changes to normalization and eigenvalue sums.

\section{Experimental Design and Target Metrics}
\label{sec:experimental_design}

\subsection{Experimental Setup and Device Selection}
\label{subsec:setup}

The channel $\mathcal{F}^\eta$ is implemented on Rigetti's Aspen-11 superconducting processor (full device specifications and coherence data in Appendix~\ref{app:device}). Different transmons were selected for the two encodings based on their coherence, readout fidelity and gate fidelity. Q45 was chosen for the qutrit implementation because of its long coherence time and high readout fidelity. Its $T_2^*$ was $31\pm4\,\mu\text{s}$, the highest on the processor, compared with $15$--$17\,\mu\text{s}$ for the two-qubit pair (Table~\ref{tab:specs}). Long coherence times help preserve the qutrit state during gate operations, while accurate readout helps distinguish its three energy levels. The two-qubit implementation used the natively coupled pair $Q_l$ (Q26) and $Q_m$ (Q27), selected for its high entangling-gate fidelities. These were $98.46\pm0.28\%$ for CZ and $98.34\pm0.38\%$ for XY. Readout fidelities were $97.15\pm0.85\%$ for $Q_l$ and $93.65\pm2.74\%$ for the ancilla $Q_m$ (Table~\ref{tab:specs}).

The qutrit implementation required additional pulse-level calibration. Virtual-$Z$ phase tracking corrected spectator-level phase errors accumulated during the $R^{01}_x(\nicefrac{\pi}{2})$ and $R^{12}_x(\nicefrac{\pi}{2})$ pulses (Appendix~\ref{app:qutrit_compilation}). A support vector machine classifier distinguished the three energy levels using their readout signatures in the I-Q plane (Appendix~\ref{app:rem}). The two-qubit implementation used the processor's standard gate compilation, with parametric XY gates and CPHASE-based CNOT gates (Appendix~\ref{app:qubit_compilation}). No additional gate calibration was performed.

\subsection{Target Metric Sampling}
\label{subsec:sampling}

The experiment tested $\mathcal{F}^{\eta}$ for 40 target metric operators with unit norm ($\|\eta\|=1$). Each metric defined a target channel, implemented and benchmarked using both encodings. The metrics were selected from two classes of PT-symmetric qubit Hamiltonians rather than sampled uniformly over the metric parameter space.

The selected metrics make two classes of unbroken PT-symmetric qubit Hamiltonians quasi-Hermitian. These classes follow the classification of Ref.~\cite{Wan13} and the discrete PT-symmetric construction of Bender \emph{et al.}~\cite{BB98, BBJ02}.
Hamiltonians in the first class are invariant under complex conjugation in the standard basis combined with the Pauli-$Z$ operation. This $Z$-symmetric class has the form
\begin{equation}
H_Z(\nu,\delta) = \epsilon\mathds{1}_2 + g\begin{pmatrix}\text{i}\nu\\ \sin\delta \\ \cos\delta\end{pmatrix}\cdot\bm\sigma,
\end{equation}
where $\nu$ is the dimensionless \emph{non-Hermiticity ratio} and $\delta$ is an angular phase parameter. For the positive-definite, unit-norm metrics used here, we restrict $0\leq\nu<1$. At $\nu=0$, $H_Z$ is Hermitian, while increasing $\nu$ increases its non-Hermitian contribution. A metric operator that makes $H_Z$ quasi-Hermitian is
\begin{equation}
\eta_Z(\nu,\delta) = \frac{1}{1+\nu}\left[\mathds{1}_2 + (0,-\nu\cos\delta,\nu\sin\delta)\cdot\bm\sigma\right].
\end{equation}
Rotating $H_Z$ gives a second class, $H_X$, invariant under complex conjugation combined with the Pauli-$X$ operation. A corresponding metric operator is
\begin{equation*}
\eta_X(\nu,\delta) = \frac{1}{1+\nu}\left[\mathds{1}_2 + (-\nu\sin\delta,-\nu\cos\delta,0)\cdot\bm\sigma\right].
\end{equation*}

Each symmetry class was sampled at 20 combinations of $\nu$ and $\delta$:
\begin{equation}
\label{eq:5nu4delta}
40\nu\in\left\{2,11,20,29,38\right\},\quad \nicefrac{4\delta}{\pi}\in\left\{-1,1,3,5\right\}.
\end{equation}
The resulting sets, $M_Z$ and $M_X$, contain 40 metrics in total. Each metric can be written as
\begin{equation*}
\eta = \frac{1}{1+\nu}\left[\mathds{1}_2 + \nu\,\hat{\bm{n}}\cdot\bm\sigma\right],
\end{equation*}
where $\hat{\bm{n}} = (\sin\theta\cos\phi, \sin\theta\sin\phi, \cos\theta)$ specifies its angular coordinates. For $M_Z$, $\hat{\bm{n}} = (0,-\cos\delta,\sin\delta)$, giving $\phi = \pm\nicefrac{\pi}{2}$ and $\theta\in\{\nicefrac{\pi}{4},\nicefrac{3\pi}{4}\}$. For $M_X$, $\hat{\bm{n}} = (-\sin\delta,-\cos\delta,0)$, giving $\theta = \nicefrac{\pi}{2}$ and $\phi\in\{-\nicefrac{\pi}{4},\nicefrac{\pi}{4},\nicefrac{3\pi}{4},\nicefrac{5\pi}{4}\}$.

The metric's minimum eigenvalue is $r=(1-\nu)/(1+\nu)$. The selected values of $\nu$ give unevenly spaced values of $r$, with more points near $r=0$. As $r$ decreases, more population can be transferred to the auxiliary level or subsystem, depending on the input state. Post-selection retains the qubit-subspace outcome in the qutrit implementation or the ancilla outcome $\ket{0}$ in the two-qubit implementation.

Appendices~\ref{app:qutrit_compilation} and~\ref{app:qubit_compilation} describe the gate compilation for the two encodings. Appendix~\ref{app:rem} covers qutrit readout and error mitigation, while Appendix~\ref{app:tomography} describes process tomography.

\section{Results and Discussion}

We implemented and benchmarked the non-unitary channel $\mathcal{F}^\eta$ for 40 target metrics using both encodings. For each metric and encoding, full quantum process tomography (QPT) yielded the experimental Choi matrix $\Lambda_\text{exp}$ from measurements of output observables for a set of input states. Appendix~\ref{app:tomography} details the reconstruction procedure.

\begin{figure}[h]
\includegraphics[width=0.84\linewidth]{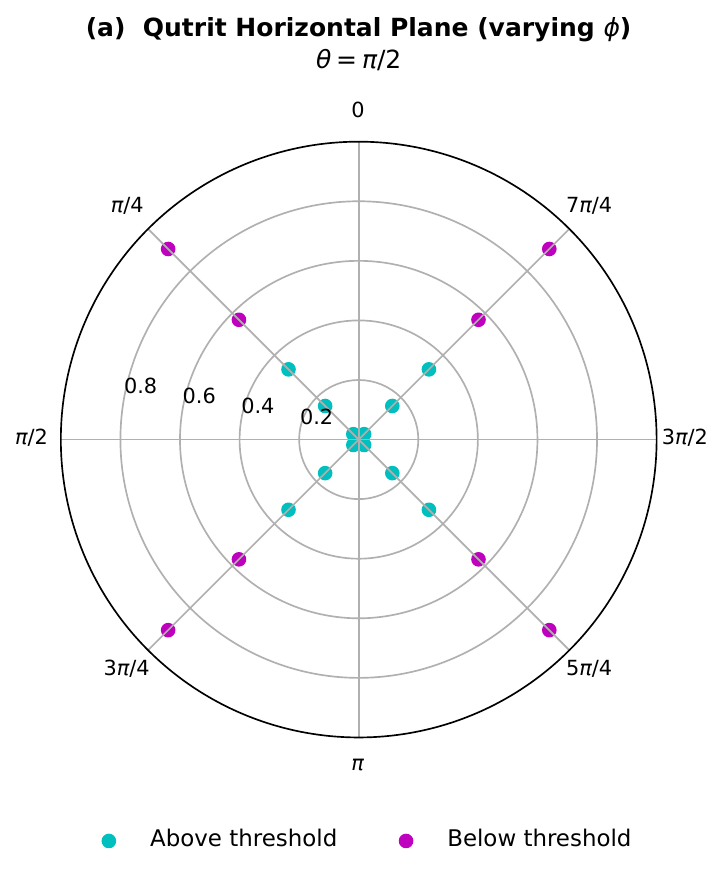}
\hfill
\includegraphics[width=0.84\linewidth]{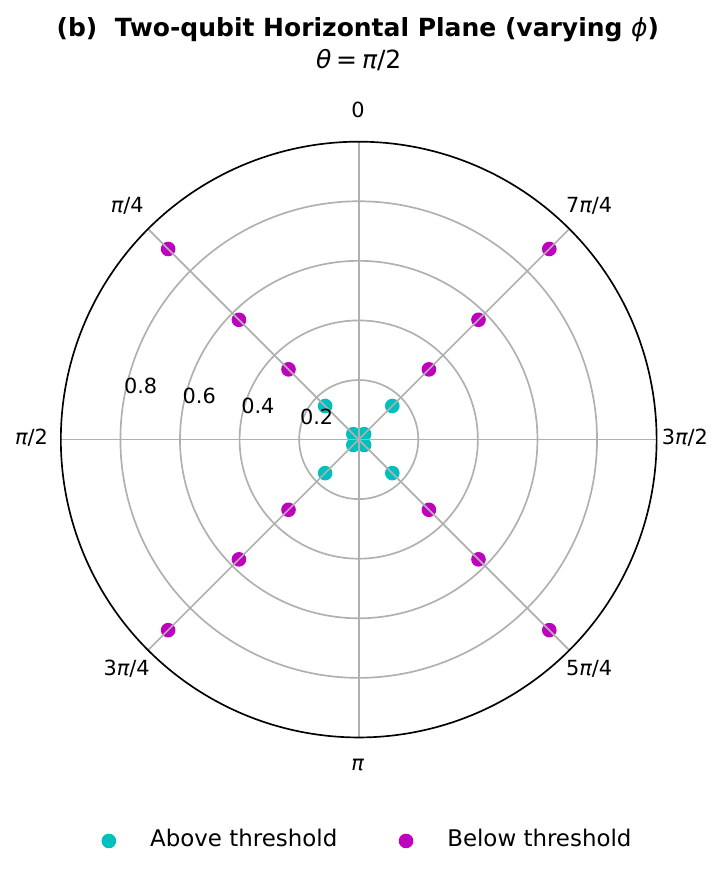}
\caption{Verification results for the 20 sampled $X$-symmetric metrics: (a) qutrit and (b) two-qubit implementations. Each point represents a metric with coordinates $(r,\theta=\nicefrac{\pi}{2},\phi)$. Cyan marks instances whose process fidelity exceeds the adversarial threshold; magenta marks those that do not. The qutrit results include readout error mitigation; the two-qubit results do not.}
\label{fig: horiz}
\end{figure}

\begin{figure}[h]
\includegraphics[width=0.84\linewidth]{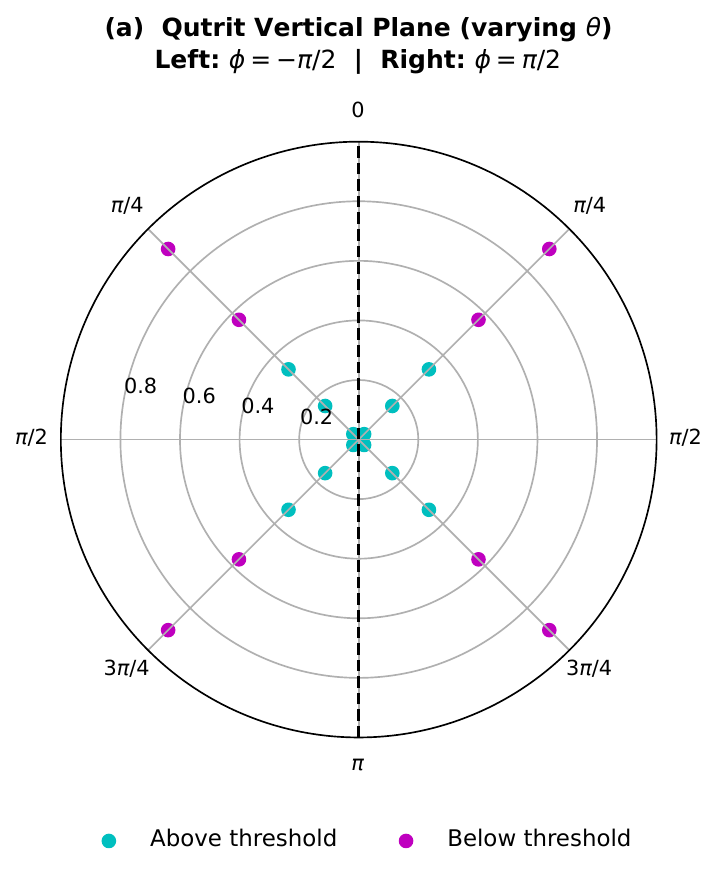}
\hfill
\includegraphics[width=0.84\linewidth]{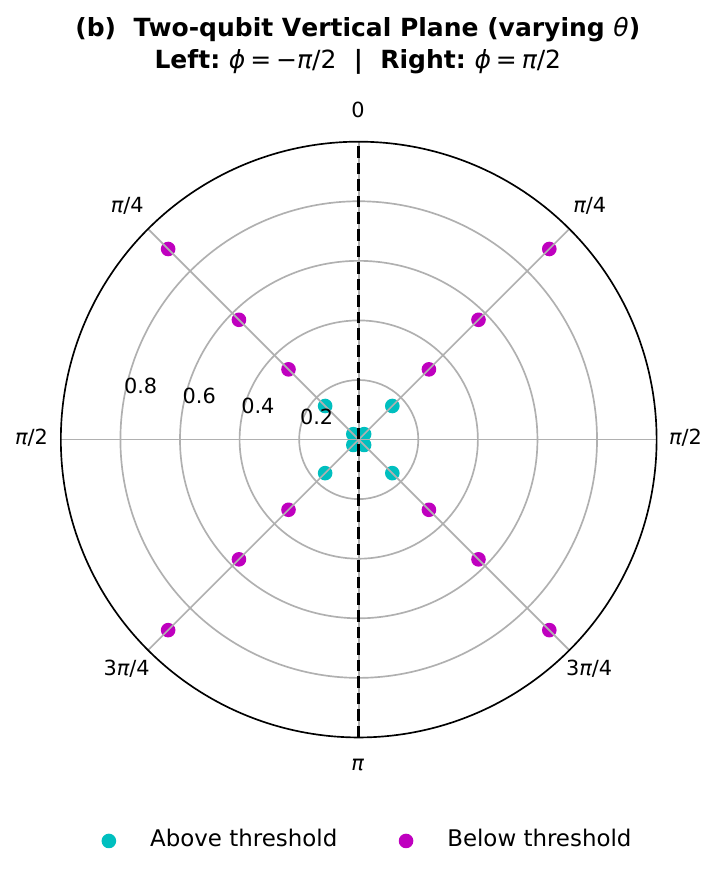}
\caption{Verification results for the 20 sampled $Z$-symmetric metrics: (a) qutrit and (b) two-qubit implementations. Each point represents a metric with coordinates $(r,\theta,\phi=\pm\nicefrac{\pi}{2})$. Cyan marks instances whose process fidelity exceeds the adversarial threshold; magenta marks those that do not. The qutrit results include readout error mitigation; the two-qubit results do not.}
\label{fig: vert}
\end{figure}

For each implementation, we calculate the process fidelity between the reconstructed and ideal Choi matrices~\cite{GLN05},
\begin{equation}
\Delta(\Lambda_{\text{exp}}, \Lambda_\eta) := \frac{\operatorname{Tr}\left[\sqrt{\sqrt{\Lambda_{\text{exp}}}\,\Lambda_\eta\sqrt{\Lambda_{\text{exp}}}}\,\right]}{\sqrt{\operatorname{Tr}[\Lambda_{\text{exp}}]}\sqrt{\operatorname{Tr}[\Lambda_\eta]}}.
\end{equation}
Throughout this work, $\Delta$ denotes the root (unsquared) fidelity between trace-normalized Choi matrices. For trace-preserving maps, the process fidelity defined in Ref.~\cite{GLN05} is $\Delta^2$. Theorem~\ref{thm:thresh} uses the root convention consistently, so an instance passes verification when its measured $\Delta$ exceeds the threshold for its metric $\eta$.
Figures~\ref{fig: horiz} and~\ref{fig: vert} show the pass/fail outcomes.
The threshold fidelity depends only on $r$. For both the two-qubit data and the error-mitigated qutrit data, all instances at a given sampled $r$ had the same pass/fail outcome. Verification applies to individual instances at the five sampled values of $r$. The largest sampled $r$ at which every instance passed therefore summarizes these measurements; it does not establish a continuous boundary for verification.

The two implementations passed verification for different fractions of the sampled metrics. Two-qubit process fidelities ranged from $0.933$ to $0.972$, with a median of $0.957$ (Appendix~\ref{app:raw_data}). All sampled metrics with $r<0.333$ passed verification, giving 16 passes out of 40 (40\%). After readout error mitigation, qutrit process fidelities ranged from $0.946$ to $0.994$, with a median of $0.981$. All sampled metrics with $r<0.569$ passed verification, giving 24 passes out of 40 (60\%).

The encodings differ in both circuit requirements and hardware: the qutrit avoids entangling gates, whereas the two-qubit implementation uses them and an ancilla readout. Their transmons also have different coherence times (Appendix~\ref{app:device}). These differences may affect fidelity, but the present measurements do not isolate their contributions.

\begin{figure*}[t]
\includegraphics[width=\textwidth]{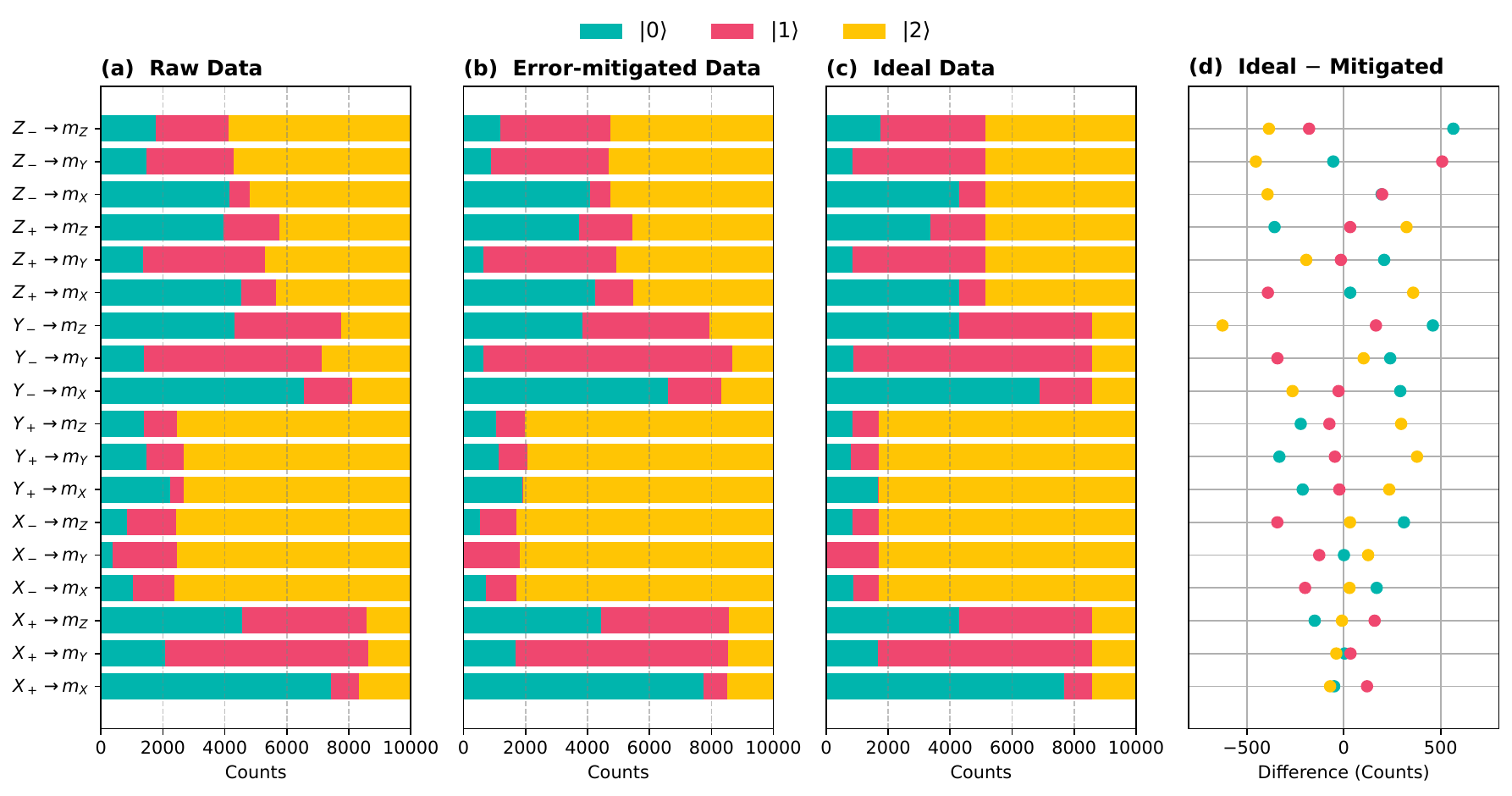}
\caption{Representative qutrit QPT data for $(r,\theta,\phi)\approx(0.026,0.50\pi,-0.25\pi)$, with 10,000 shots per input-state and measurement-basis combination. Each row corresponds to one of the 18 tomography settings (Appendix~\ref{app:tomography}). Columns show (a) raw data, (b) readout-error-mitigated data (Appendix~\ref{app:rem}), (c) ideal outputs and (d) ideal minus mitigated outputs. The row labels $S_{\mathrm{in}}\rightarrow m_{\mathrm{out}}$ identify the prepared state $S_{\mathrm{in}}\in\{Z_\pm,Y_\pm,X_\pm\}$ in the qubit subspace and the measured observable $m_{\mathrm{out}}\in\{m_Z,m_Y,m_X\}$.}
\label{fig: qpt-table}
\end{figure*}

Readout error mitigation (REM) was applied only to the qutrit data, because the two-qubit measurements preceded adoption of the procedure (Appendix~\ref{app:device}). For the qutrit, decay from $\ket{2}$ during readout increases errors in the three-state classification, so we calibrated a $3\times3$ assignment matrix before each set of metric parameters (Appendix~\ref{app:rem}). Figure~\ref{fig: qpt-table} shows a representative qutrit QPT dataset before and after this correction. REM corrects classical readout errors without improving the coherent dynamics. The unequal treatment therefore limits comparisons of the encodings; equivalent mitigation could also benefit the two-qubit data, particularly because post-selection depends on the ancilla $Q_m$, whose readout fidelity was $93.65\%$ (Table~\ref{tab:specs}).

Without REM, the qutrit implementation passes 14 of 40 instances and has a median fidelity of $0.938$, compared with 16 passes for the unmitigated two-qubit implementation (Appendix~\ref{app:raw_data}). Mitigation adds ten qutrit passes: the eight instances at $r=0.333$ and instances 7 and 8 at $r=0.159$. Thus, the raw qutrit data give mixed outcomes at $r=0.159$, where those two instances fail and the other six pass. Since certification requires an accurate fidelity estimate (Appendix~\ref{app:threshold}), these additional passes depend on how accurately the assignment-matrix model describes the readout errors.

\section{Conclusion and Outlook}
This experiment demonstrates the first direct simulation of a Hilbert-space IP change on a superconducting processor. Building on Ref.~\cite{KAS22}, we derived a process-fidelity threshold for the IP-change channel and tested it with both encodings. This single-Kraus channel maps pure inputs to pure conditional outputs and is trace-non-increasing; the survival probability depends on the input. A process fidelity above this threshold rules out mixtures of unitary operations within the qubit subspace, even with random discarding of runs. Under the benchmark's assumptions, exceeding the threshold certifies the use of higher-dimensional coherence.

We applied this benchmark to two superconducting encodings of the channel: a single qutrit and a two-qubit Stinespring dilation. For both encodings, all instances at a given sampled $r$ had the same pass/fail outcome after readout error mitigation of the qutrit data. The mitigated qutrit data passed verification over a larger range of sampled $r$ than the unmitigated two-qubit data. Without mitigation, the two-qubit implementation passed 16 of 40 instances, compared with 14 for the qutrit. The unequal use of mitigation limits this comparison, and several hardware differences may also affect the results. The qutrit implementation used the transmon with the highest available $T_2^*$ (Sec.~\ref{subsec:setup}). The two-qubit pair was selected primarily for its entangling-gate fidelities. Its performance also depended on coherence times and the readout fidelity of the ancilla $Q_m$, whose outcome determines post-selection (Table~\ref{tab:specs}). Applying equivalent readout mitigation to both encodings would enable a more direct comparison.

Although the experiment tested metrics from two symmetry classes, the constructions apply more broadly to positive-definite qubit metrics with eigenvalues $1$ and $r\in(0,1]$. The two-qubit dilation also accommodates general $0<\eta\leq\mathds{1}_2$, while the displayed qutrit construction assumes unit norm. A next step is to use IP changes to simulate unbroken PT-symmetric evolution and adapt the benchmark to test those simulations. The adversarial threshold extends to arbitrary dimension (Appendix~\ref{app:threshold}), but full quantum process tomography makes experimental verification costly. Efficient implementations of IP changes may also help develop quantum algorithms involving non-Hermitian Hamiltonians~\cite{AK24,ZZHY25}.

\section*{Data Availability}
The raw and error-mitigated datasets are publicly available on Zenodo at \url{https://doi.org/10.5281/zenodo.24176060}. The repository also contains the code for quantum process tomography, readout error mitigation and figure generation.

\begin{acknowledgments}
This work is supported by the U.S. Department of Energy, Office of Science, National Quantum Information Science Research Centers, Superconducting Quantum Materials and Systems Center
(SQMS) under contract No. DE-AC02-07CH11359. We thank Matthew Reagor and Rigetti Computing for access to the Aspen-11 processor and for technical support.The authors used Claude (Anthropic), Gemini (Google) and ChatGPT (OpenAI) to improve the manuscript's language and readability. Gemini also assisted with figure formatting. Claude's assistance with the analysis code is
described in Appendix~\ref{app:tomography}. The authors produced all data and scientific content. They reviewed all AI-assisted content and take full responsibility for the work.
\end{acknowledgments}

\appendix
\section{Qutrit Unitary Construction and Compilation}
\label{app:qutrit_compilation}
This appendix proves that the gate sequence in Eq.~\eqref{eq:3circuit} realizes $U_3^\eta$ defined in Sec.~\ref{subsec:implementation}. It also describes how to compile the sequence into native pulses.

\subsection{Proof of Phase-Locked Parameters}
To show that the gate sequence in Eq.~\eqref{eq:3circuit} realizes Eq.~\eqref{eq:U3eta} with $\xi=\varphi=0$, we first write $R^{12}_y(-\gamma)$ as
\begin{equation}
R^{12}_y(-\gamma) = \begin{pmatrix} 1 & 0 & 0 \\ 0 & \sqrt{r} & \sqrt{1-r} \\ 
0 & -\sqrt{1-r} & \sqrt{r} \end{pmatrix} \equiv \begin{pmatrix} D & \bm{w} \\ -\bm{w}^\top & \sqrt{r} \end{pmatrix},
\end{equation}
where $D = \operatorname{diag}(1, \sqrt{r})$ and $\bm{w} = (0, \sqrt{1-r})^\top$.

Define $V=R^{01}_z(\phi)R^{01}_y(\theta)$, so that $\eta^{1/2}=VDV^\dagger$. The gate sequence then gives
\begin{equation*}
\tilde{U}_3^\eta = (V\oplus 1)\,R^{12}_y(-\gamma)\,(V^\dagger\oplus 1).
\end{equation*}
Multiplying the blocks gives
\begin{equation}
\tilde{U}_3^\eta = \begin{pmatrix} V D V^\dagger & V \bm{w} \\ -\bm{w}^\top V^\dagger & \sqrt{r} \end{pmatrix}.
\end{equation}

Comparing the blocks with $U_3^\eta(\xi,\varphi)$ gives:
\begin{enumerate}
    \item \textbf{Top-left block:} $VDV^\dagger=\eta^{1/2}$.
    \item \textbf{Top-right block:} $V\bm{w}=\sqrt{1-r}\,\bm{v}_r$, where $\bm{v}_r=V\ket{1}$. This matches $\bm{u}=\text{e}^{\text{i}\varphi}\sqrt{1-r}\,\bm{v}_r$ with $\varphi=0$.
    \item \textbf{Bottom-left block:} $-\bm{w}^\top V^\dagger=-\sqrt{1-r}\,\bm{v}_r^\dagger$. With $\bm{u}=\sqrt{1-r}\,\bm{v}_r$, this matches $-\text{e}^{\text{i}\xi}\bm{u}^\dagger$ with $\xi=0$.
\end{enumerate}
The gate sequence therefore realizes $U_3^\eta$ with $\xi=\varphi=0$ and reproduces the target channel after projection.

\subsection{Pulse-level Native Decomposition}

We compiled the circuits into native instructions for the Aspen-11 processor~\cite{valeryDynamicalSweetSpot2022a}. Quil-T, Rigetti's pulse-level control assembly language, provided control for calibration and execution~\cite{quilt}.

The decomposition of $U_3^\eta$ uses subspace rotations $R_y$ and $R_z$. The subspace rotation $R^{ij}_n(\theta)$ applies $\exp(-\text{i}\theta\sigma_n/2)$ within the subspace spanned by $\{\ket{i},\ket{j}\}$. It acts as the identity on the spectator level.

The native gate set combines fixed-amplitude microwave pulses $R^{01}_x(\nicefrac{\pi}{2})$ and $R^{12}_x(\nicefrac{\pi}{2})$ with virtual $Z$-gates tracked in software~\cite{mckayEfficientGatesQuantum2017,gossHighfidelityQutritEntangling2022, fischerAncillafreeImplementationGeneralized2022,caoEmulatingTwoQubits2023}. The virtual phase shifts are defined as:
\begin{equation}
RZ_{01}\left(\zeta\right) = 
\begin{pmatrix}
\text{e}^{\text{i}\zeta} & 0 & 0\\
0 & 1 & 0\\
0 & 0 & 1
\end{pmatrix}, \quad
RZ_{12}\left(\zeta\right) = 
\begin{pmatrix}
1 & 0 & 0\\
0 & 1 & 0\\
0 & 0 & \text{e}^{\text{i}\zeta}
\end{pmatrix}.
\end{equation}
Driving a transition in a multilevel transmon can also shift the phase of the spectator level. We measured spectator-level phase errors during the $R^{01}_x(\nicefrac{\pi}{2})$ and $R^{12}_x(\nicefrac{\pi}{2})$ pulses using the method of Ref.~\cite{morvanQutritRandomizedBenchmarking2021}. Calibrated virtual phase updates on both transitions corrected these errors. With these phase corrections, the native-gate decompositions used to construct $U_3^\eta$ are given below. Each equality holds up to a global phase, and the rightmost operator acts first.
\begin{align}
R^{01}_z(\phi) =& RZ_{01}(-\phi)RZ_{12}(-\nicefrac{\phi}{2}), \\
R^{01}_y(\theta) =& R^{01}_x(\nicefrac{\pi}{2})RZ_{01}(-\pi+\theta)RZ_{12}(-(\pi-\theta)/2) \nonumber \\ 
                  &\times R^{01}_x(\nicefrac{\pi}{2})RZ_{01}(\pi)RZ_{12}(\nicefrac{\pi}{2}), \\
R^{12}_y(-\gamma) =& R^{12}_x(\nicefrac{\pi}{2})RZ_{01}((\pi+\gamma)/2)RZ_{12}(\pi+\gamma) \nonumber \\
                   &\times R^{12}_x(\nicefrac{\pi}{2})RZ_{01}(-\nicefrac{\pi}{2})RZ_{12}(-\pi).
\end{align}
These decompositions implement the required rotations and compensate spectator-level phases before the final projection $P$.

\section{Two-Qubit Stinespring Dilation and Native Gates}
\label{app:qubit_compilation}
This appendix describes how to compile the Stinespring dilation unitary $U_{2\times2}^\eta$ from Sec.~\ref{subsec:implementation} into native gates.

The processor provides two native entangling gates: controlled-phase (CPHASE) and parametric XY~\cite{KTP+20,PBE22}. The CNOT in Eq.~\eqref{eq:2circuit} uses a native $\text{CPHASE}(\pi)$ gate preceded and followed by single-qubit rotations.

The native parametric gate ${\rm XY}(\alpha,\vartheta)$ exchanges an excitation between the capacitively coupled transmons. In the computational basis $\{\ket{00},\ket{01},\ket{10},\ket{11}\}$, its matrix is
\begin{equation}
\label{eq:XY}
{\rm XY}(\alpha,\vartheta) = \begin{pmatrix}
1 & 0 & 0 & 0 \\
0 & \cos(\nicefrac{\vartheta}{2}) & \text{i}\sin(\nicefrac{\vartheta}{2}) \text{e}^{\text{i}\alpha} & 0 \\
0 & \text{i}\sin(\nicefrac{\vartheta}{2}) \text{e}^{-\text{i}\alpha} & \cos(\nicefrac{\vartheta}{2}) & 0 \\
0 & 0 & 0 & 1
\end{pmatrix},
\end{equation}
where $\vartheta\in[0,\pi]$ is the rotation angle and $\alpha\in[0,2\pi)$ is the phase.
In Eq.~\eqref{eq:2circuit}, ${\rm XY}(\vartheta)$ denotes ${\rm XY}(0,\vartheta)$. The chosen angle, $\vartheta=\gamma=2\cos^{-1}(\sqrt{r})$, lies in $[0,\pi]$ for $r\in[0,1]$. Define $V=R_z(\phi)R_y(\theta)$, as in the qutrit construction (Appendix~\ref{app:qutrit_compilation}). The initial rotation $V^\dagger$ maps the eigenvectors of $\eta$ with eigenvalues $1$ and $r$ to $\ket{0}$ and $\ket{1}$ of $Q_l$, respectively. In this eigenbasis, the sequence $\mathrm{CNOT}\,{\rm XY}(\gamma)\,\mathrm{CNOT}$ uses $Q_m$ as the control for both CNOT gates. It leaves $\ket{0}_l\ket{0}_m$ unchanged and maps
\begin{equation}
\ket{1}_l\ket{0}_m \mapsto \sqrt{r}\,\ket{1}_l\ket{0}_m + \text{i}\sqrt{1-r}\,\ket{1}_l\ket{1}_m.
\end{equation}
Post-selection on the ancilla outcome $\ket{0}$ applies $\operatorname{diag}(1,\sqrt{r})$ to $Q_l$. Including the surrounding rotations gives $V\operatorname{diag}(1,\sqrt{r})V^\dagger=\eta^{1/2}$, as required by Eq.~\eqref{eq:unitary_projection}. The phase $\alpha$ affects only the amplitude transferred to $\ket{1}_m$. Post-selection discards this outcome, so $\alpha$ does not affect the channel. Setting $\alpha=0$ reproduces the block form of $U^\eta_{2\times2}$ given in Sec.~\ref{subsec:implementation}. As in the qutrit implementation, each $R^{01}_y$ rotation uses $R^{01}_x(\nicefrac{\pi}{2})$ pulses and virtual $Z$-gates. This compilation aims to reduce microwave crosstalk and gate duration.

\section{Qutrit Dispersive Readout and Error Mitigation}
\label{app:rem}
Characterizing the qutrit channel requires distinguishing the lowest three transmon energy levels in single-shot measurements. A capacitively coupled resonator provides dispersive readout of the transmon. The effective Hamiltonian is~\cite{tavis1968exact, blais2004cavity, blais2021circuit}
\begin{equation}
H = \omega_c a^{\dagger} a + \omega_q b^{\dagger} b - \frac{E_C}{2} b^{\dagger} b^{\dagger} b b + \sum_{j=0}^{2} (\Lambda_j + \chi_j a^{\dagger} a )\ket{j}\bra{j},
\end{equation}
where $\omega_c$ is the cavity frequency, $b^\dagger$ is the transmon creation operator, and $E_C$ is the charging energy. The dispersive shift $\chi_j=\chi_{j-1,j}-\chi_{j,j+1}$ makes the resonator frequency depend on the transmon energy level. We selected the readout frequency and amplitude to distinguish the three states in the I-Q plane. A support vector machine (SVM) classifier assigned the measured microwave signals to these states~\cite{boser1992training}. The classifier used \texttt{scikit-learn}~\cite{scikit-learn} and was trained on readout signals from preparations of $\ket{0}$, $\ket{1}$ and $\ket{2}$.

State-preparation and measurement errors affect this classification. In particular, higher excited states can decay during readout, causing their I-Q signatures to overlap with those of lower states. A $3\times3$ assignment matrix corrects readout misassignment errors before process-fidelity estimation and adversarial verification~\cite{Maciejewski2020mitigationofreadout}.

To construct the assignment matrix $\mathbf{M}$, we prepared each basis state $\ket{j}\in\{\ket{0},\ket{1},\ket{2}\}$ and recorded the frequencies of the readout labels $i\in\{0,1,2\}$. Each element gives the probability of assigning label $i$ after preparing state $\ket{j}$:
\begin{equation}
\label{eq:assignment_matrix}
M_{ij} = P(i|j).
\end{equation}

Linear inversion converts the measured probability vector $\vec{P}_{\text{raw}}$ into the corrected vector $\vec{P}_{\text{mit}}$:
\begin{equation}
\label{eq:rem_inversion}
\vec{P}_{\text{mit}} = \mathbf{M}^{-1} \vec{P}_{\text{raw}}.
\end{equation}
Statistical noise can make some entries of the inverted vector negative. We set these entries to zero and normalize the remaining entries so that $\sum_i(P_{\text{mit}})_i=1$. We recalibrated $\mathbf{M}$ before each new set of metric parameters to account for slow thermal fluctuations in the readout resonator.

\section{Tomography and Process Matrix Reconstruction}
\label{app:tomography}
We characterized the implemented IP change using process tomography on the qubit subspace~\cite{CN97, NC00}. The process matrix $\chi$ represents the quantum channel. For a single-qubit channel with Kraus operators $\{E_i\}$, choose the Pauli basis $\Sigma:=\{\mathds{1}_2,X,Y,Z\}$. The $4\times4$ process matrix has elements
\begin{equation*}
\chi_{mn}:=\sum_i\operatorname{Tr}(\Sigma_m E_i)\operatorname{Tr}^*(\Sigma_n E_i),
\end{equation*}
where $\operatorname{Tr}^*(\bullet):=\overline{\operatorname{Tr}(\bullet)}$ denotes the complex conjugate of the trace. The resulting matrix is Hermitian: $\chi_{nm}=\overline{\chi_{mn}}$. For $\mathcal{F}^\eta$, the ideal process matrix is
\begin{equation}
\chi_{mn}^\eta = \operatorname{Tr}(\Sigma_m \eta^{1/2}) \operatorname{Tr}^*(\Sigma_n \eta^{1/2}).
\end{equation}

To estimate $\chi_{\text{exp}}$, we prepared an informationally complete set of six single-qubit input states:
\begin{equation}
\label{eq:sixqubitstates}
\ket{0}, \ket{1}, \ket{\pm}, \ket{\pm \text{i}}.
\end{equation}
Each input underwent $U^\eta_3$ or $U^\eta_{2\times2}$, followed by measurement in the $X$, $Y$ and $Z$ bases. Each shot produced one of three recorded outcomes: $0$, $1$ or $\emptyset$. The outcome $\emptyset$ denotes failure of the post-selection condition.

Each of the 18 input-state and measurement-basis combinations used $N$ shots to estimate the corresponding expectation value. With $N=10{,}000$ shots per combination, the statistical uncertainty scales as $1/\sqrt{N}=0.01$~\cite{STM11}.

Linear inversion of the measurement data gives the estimate $\chi_{\text{exp}}$. Let $O_k(i,j)$ denote the fractional count of outcome $k \in \{0, 1, \emptyset\}$ for input state $i \in \{1, \dots, 6\}$ and Pauli measurement basis $j \in \{1, 2, 3\}$. The Pauli expectation values are
\begin{equation*}
\tilde{O}(i,j):=O_0(i,j)-O_1(i,j),\qquad j=1,2,3.
\end{equation*}
A fourth entry records the survival probability $\operatorname{Tr}[\mathcal{F}^\eta(\rho_i)]$, which is the identity component of the output. Averaging the survival estimates from the three measurement bases gives
\begin{align*}
\tilde{O}(i,4)&:=\frac{1}{3}\sum_{j=1}^3[1-O_\emptyset(i,j)]\\
&=\frac{1}{3}\sum_{j=1}^3[O_0(i,j)+O_1(i,j)].
\end{align*}
The resulting $6\times4$ array contains three Pauli expectation values and one survival probability for each input state. Flattening this array gives the 24-component data vector $\vec{\lambda}$.

The analysis code now estimates the survival probability by averaging over all three measurement bases, replacing the previous X-basis-only estimate. The authors specified this change and used Claude Fable 5.2 (Anthropic) to assist with its implementation. They reviewed the revised code and checked that restricting the calculation to the X basis reproduces the previous results. The updated code is available in the Zenodo repository~\cite{zenodo_data}.

The measurement matrix $\beta\in\mathbb{C}^{24\times16}$ relates the data vector to the vectorized process matrix through $\vec{\lambda}\approx\beta\vec{\chi}$. Each row of $\beta$ corresponds to a data entry $(i,j)$ and contains the vectorized operator $\rho_i\otimes P_j$. Here, $\rho_i$ is the input-state density matrix, $P_1=X$, $P_2=Y$, $P_3=Z$ and $P_4=\mathds{1}_2$. Applying the Moore--Penrose pseudoinverse of $\beta$ gives $\vec{\tilde{\chi}}=\beta^+\vec{\lambda}$. Reshaping this vector gives the $4\times4$ unconstrained estimate $\tilde{\chi}$.

Finite sampling noise can give the reconstructed Choi matrix negative eigenvalues or produce a map that increases the trace. After transforming $\tilde{\chi}$ into the Choi representation $\tilde{\Lambda}$, we use Forest-Benchmarking's \texttt{proj\_choi\_to\_physical} routine with \texttt{make\_trace\_preserving=False} to enforce complete positivity and the trace-non-increasing condition. The routine uses Dykstra's alternating-projection algorithm, following Knee \emph{et al.}~\cite{knee2018quantum}, to approximate the least-squares projection
\begin{equation}
\Lambda_{\text{exp}} = \operatorname*{arg\,min}_{\substack{\Lambda \ge 0\\ \operatorname{Tr}_{\rm out}[\Lambda] \le \mathds{1}_2}} \left\| \Lambda - \tilde{\Lambda} \right\|_{\rm F}^2,
\end{equation}
where $\|\cdot\|_{\rm F}$ is the Frobenius norm, $\Lambda\geq0$ denotes a Hermitian positive semidefinite matrix and $\operatorname{Tr}_{\rm out}$ is the partial trace over the output system. The process matrix $\chi_{\text{exp}}$ is $\Lambda_{\text{exp}}$ expressed in the Pauli basis. The fidelity of $\Lambda_{\text{exp}}$ with the ideal matrix $\Lambda_\eta$ gives the experimental process fidelity.

\section{Adversarial Threshold Derivation}
\label{app:threshold}
The adversarial benchmark bounds the process fidelity achievable by a restricted class of operations within the qubit subspace. It extends the tomographic verification scheme of Ref.~\cite{KAS22}. The target IP-change channel $\mathcal{F}^\eta$ can reduce the trace of the input state. The adversary can apply mixtures of unitary operations within the qubit subspace and randomly discard runs. The model excludes access to higher energy levels or an entangled ancilla.

The adversarial model allows mixtures of unitary operations and random discarding within $\mathcal{H}_2$. It does not cover all trace-non-increasing completely positive maps on $\mathcal{H}_2$. The threshold below applies only to this restricted class. Reference~\cite{KAS22} derives its verification threshold in terms of the induced Schatten $(1\to1)$-norm distance between the ideal and adversarial operations. Theorem~\ref{thm:thresh}, restated below, gives a threshold in terms of process fidelity~\cite{GLN05}. This is the quantity calculated from our tomographic reconstruction (Appendix~\ref{app:tomography}) and used throughout this work.

The adversarial operation is
\begin{equation}\label{eq:disG}
   {\mathcal{F}}_{\eta}^{\rm dis}(\rho):\mathcal{B}\left(\mathcal{H}_2\right)\to \mathcal{B}\left(\mathcal{H}_2\right):\rho\mapsto \sum_{j=1}^Mq_jU_j\rho U_j^\dagger, 
\end{equation}
where each $U_j$ is unitary and $q_j\geq0$ is the probability of applying $U_j$ and retaining the run. The remaining probability, $1-\sum_jq_j\geq0$, corresponds to discarding the run.

The metric eigenvalues $\lambda_1\geq\lambda_2$ determine the threshold $D_{\rm th}^{(2)}(\eta)$. The following theorem bounds the process fidelity between the ideal channel $\mathcal{F}^{\eta}$ and any adversarial operation $\mathcal{F}_{\eta}^{\rm dis}$ by this threshold.

\begin{restatethm}{\ref{thm:thresh}}
Let $\eta$ be a metric operator on $\mathcal{H}_2$ satisfying $\eta\leq\mathds{1}_2$, with eigenvalues $\lambda_1\geq\lambda_2$. Define
\begin{align*}
\mathcal{F}^{\eta}(\rho)&=\eta^{1/2}\rho\eta^{1/2},\\
\mathcal{F}_{\eta}^{\rm dis}(\rho)&=\sum_{j=1}^{M}q_jU_j\rho U_j^\dagger,
\end{align*}
where each $U_j$ is unitary, $q_j\geq0$, and $0<\sum_jq_j\leq1$. Then
\begin{equation}\label{eq:accrej2}
  \Delta\left(\mathcal{F}_{\eta}^{\rm dis},\mathcal{F}^{\eta}\right) \le  D^{(2)}_{\rm th}(\eta) :=  \frac{\sqrt{\lambda_1}+\sqrt{\lambda_2}}{\sqrt{2(\lambda_1+\lambda_2)}}.
\end{equation}
\end{restatethm}

\begin{proof}
The ideal channel $\mathcal{F}^{\eta}$ has Choi matrix 
\begin{align}\label{eq:idealChoi}
\Lambda_{{\eta}} &=  2 \left(\mathcal{F}^{\eta}\otimes \mathcal{I}_2\right)\ket{\Omega}\bra{\Omega} \nonumber \\
&= 
2\left(\eta^{1/2}\otimes\mathds{1}_2\right)\ket{\Omega}\bra{\Omega} \left(\eta^{1/2}\otimes\mathds{1}_2\right), 
\end{align}
where $\ket{\Omega} := \frac{1}{\sqrt{2}}\left(\ket{00}+\ket{11}\right)$ is the normalized maximally entangled (Bell) state on $\mathcal{H}_2\otimes\mathcal{H}_2$. The adversarial operation $\mathcal{F}^{\rm dis}_{\eta}$ has Choi matrix
\begin{align}\label{eq:disChoi}
\Lambda_{\mathcal{F}^{\rm dis}_{\eta}} &= 2 \left(\mathcal{F}^{\rm dis}_{\eta}\otimes \mathcal{I}_2\right)\ket{\Omega}\bra{\Omega} \nonumber \\
&= 2\sum_{j=1}^M q_j \left(U_j\otimes \mathds{1}_2\right) \ket{\Omega}\bra{\Omega}\left(U_j^\dagger\otimes \mathds{1}_2\right).
\end{align}
The process fidelity between $\mathcal{F}^{\eta}$ and $\mathcal{F}^{\rm dis}_{\eta}$~\cite{GLN05} is
\begin{equation}\label{eq:procfidelity}
  \Delta\left(\mathcal{F}^{\eta},\mathcal{F}^{\rm dis}_{\eta}\right)=  \frac{\operatorname{Tr}\left[\sqrt{\sqrt{\Lambda_{{\eta}}} {\Lambda}_{\mathcal{F}^{\rm dis}_{\eta}} \sqrt{\Lambda_{{\eta}}}}\right]}{\sqrt{\operatorname{Tr}\left[\Lambda_{{\eta}}\right]} \sqrt{\operatorname{Tr}\left[\Lambda_{\mathcal{F}^{\rm dis}_{\eta}}\right]}}.
\end{equation}
The normalization $\braket{\Omega|\Omega}=1$ gives
\begin{align*}
\operatorname{Tr}[\Lambda_\eta]&=\operatorname{Tr}[\eta],\\
\operatorname{Tr}[\Lambda_{\mathcal{F}^{\rm dis}_\eta}]&=2\sum_{j=1}^{M}q_j.
\end{align*}

A projector is its own positive square root. Since $\Lambda_{{\eta}}$ is proportional to a rank-one projector,
\begin{equation}\label{eq:choiidealsq}
 \sqrt{\Lambda_{{\eta}}} = \sqrt{\operatorname{Tr}[\Lambda_{{\eta}}]}  \sqrt{\frac{\Lambda_{{\eta}}}{\operatorname{Tr}[\Lambda_{{\eta}}]}} =  
 \frac{\Lambda_{{\eta}}}{\sqrt{\operatorname{Tr}[\Lambda_{{\eta}}]}}.
\end{equation}
Therefore,
\begin{equation*}
\sqrt{\Lambda_{{\eta}}} {\Lambda}_{\mathcal{F}^{\rm dis}_{\eta}} \sqrt{\Lambda_{{\eta}}} = \frac{1}{{\operatorname{Tr}[\Lambda_{{\eta}}]}}\Lambda_{{\eta}} {\Lambda}_{\mathcal{F}^{\rm dis}_{\eta}}\Lambda_{{\eta}}.
\end{equation*}

Combining Eqs.~\eqref{eq:idealChoi} and~\eqref{eq:disChoi} gives 
\begin{align}
\Lambda_{{\eta}} {\Lambda}_{\mathcal{F}^{\rm dis}_{\eta}}\Lambda_{{\eta}} &= 8\sum_{j=1}^M q_j \left|\bra{\Omega}\left(\eta^{1/2}U_j\otimes \mathds{1}_2\right)\ket{\Omega}\right|^2 \nonumber \\
&\quad \times \left(\eta^{1/2}\otimes\mathds{1}_2\right)\ket{\Omega}\bra{\Omega} \left(\eta^{1/2}\otimes\mathds{1}_2\right).
\end{align}
The definition of $\ket{\Omega}$ gives
\begin{equation*}
\left|\bra{\Omega}(\eta^{1/2}U_j\otimes\mathds{1}_2)\ket{\Omega}\right|^2=\frac14\left|\operatorname{Tr}[\eta^{1/2}U_j]\right|^2.
\end{equation*}
The product $\Lambda_\eta\Lambda_{\mathcal{F}^{\rm dis}_\eta}\Lambda_\eta$ is therefore proportional to a rank-one projector. Its square root is
\begin{align}\label{eq:numer2}
    \sqrt{\Lambda_{{\eta}} {\Lambda}_{\mathcal{F}^{\rm dis}_{\eta}}\Lambda_{{\eta}}} &= \frac{2\sqrt{\sum_{j=1}^M q_j \left|\operatorname{Tr}\left[\eta^{1/2} U_j\right]\right|^2}}{\sqrt{\operatorname{Tr}\left[\eta\right]}} \nonumber \\
    &\quad \times \left(\eta^{1/2}\otimes\mathds{1}_2\right)\ket{\Omega}\bra{\Omega} \left(\eta^{1/2}\otimes\mathds{1}_2\right).
\end{align}
The numerator of the process fidelity is
\begin{equation}
    \operatorname{Tr}\left[ \sqrt{\sqrt{\Lambda_{{\eta}}} {\Lambda}_{\mathcal{F}^{\rm dis}_{\eta}} \sqrt{\Lambda_{{\eta}}}}\right] ={\sqrt{\sum_{j=1}^M q_j \left|\operatorname{Tr}\left[\eta^{1/2} U_j\right]\right|^2}}.
\end{equation}
Substituting the numerator and trace factors gives 
\begin{equation}
 \Delta\left(\mathcal{F}^{\eta},\mathcal{F}^{\rm dis}_{\eta}\right) = \frac{\sqrt{\sum_{j=1}^M q_j \left|\operatorname{Tr}\left[\eta^{1/2} U_j\right]\right|^2}}{\sqrt{\operatorname{Tr}\left[\eta\right]}\sqrt{2\sum_{j=1}^M q_j}}.
\end{equation}
Since $\eta^{1/2}$ is positive definite, von Neumann's trace inequality gives
\begin{equation*}
\left|\operatorname{Tr}[\eta^{1/2}U_j]\right|\leq\operatorname{Tr}[\eta^{1/2}].
\end{equation*}
Substituting this inequality gives the bound
\begin{equation}
   \Delta\left(\mathcal{F}^{\eta},\mathcal{F}^{\rm dis}_{\eta}\right)\le  
   \frac{\sqrt{\sum_{j=1}^M q_j \operatorname{Tr}\left[\eta^{1/2}\right]^2}}{\sqrt{\operatorname{Tr}\left[\eta\right]}\sqrt{2\sum_{j=1}^M q_j}} = \frac{\sqrt{\lambda_1}+\sqrt{\lambda_2}}{\sqrt{2(\lambda_1+\lambda_2)}},
\end{equation}
where $\lambda_1,\lambda_2$ are the eigenvalues of $\eta$.  
\end{proof}

\begin{remark}
The bound in Eq.~\eqref{eq:accrej2} is the qubit case of the general threshold
\begin{equation}
    D^{(d)}_{\rm th}(\eta) 
    := \frac{\operatorname{Tr}[\eta^{1/2}]}{\sqrt{d \cdot \operatorname{Tr}[\eta]}}
     = \frac{\sum_{k=1}^{d}\sqrt{\lambda_k}}
            {\sqrt{d\sum_{k=1}^{d}\lambda_k}},
\end{equation}
where $\lambda_1,\ldots,\lambda_d$ are the eigenvalues of $\eta$. The same argument extends to dimension $d$. The ideal Choi matrix remains rank one, and von Neumann's trace inequality still applies. Replacing the qubit normalization factors with their $d$-dimensional counterparts gives the threshold above. The experiment focuses on $d=2$, the simplest nontrivial case.
\end{remark}

A true process fidelity above $\Delta_\text{th}^\eta$ rules out every strategy in the defined adversarial class. Under the benchmark's assumptions, this certifies the use of higher-dimensional coherence. Experimental certification requires accounting for uncertainty in the estimated fidelity. Tomographic estimates can retain systematic errors, including residual SPAM errors after mitigation. These errors can overestimate the fidelity and produce an apparent threshold crossing even when the true fidelity remains below $\Delta_\text{th}^\eta$.

\section{Device Specifications and Coherence}
\label{app:device}
The experiments used Rigetti's Aspen-11 processor. The two-qubit experiment ran on October 5, 2022, using the natively coupled transmons $Q_l$ (Q26) and ancilla $Q_m$ (Q27). The qutrit experiment used Q45 on October 10--11, 2022.

Table~\ref{tab:specs} lists the hardware parameters, coherence times and median gate fidelities measured during these experiments.

\begin{table}[htbp]
\caption{\label{tab:specs} Hardware specifications of the Aspen-11 transmons used in the IP-change experiments. The uncertainties in coherence times and fidelities are standard deviations measured during each experiment.}
\addtolength{\tabcolsep}{-1pt}
\begin{ruledtabular}
\begin{tabular}{lccc}
Parameter & Qutrit (Q45) & $Q_l$ (Q26) & $Q_m$ (Q27) \\
\hline
$\omega_{01}/2\pi$ (GHz) & 3.595              & 5.131              & 3.895 \\
Anharm. (MHz)            & 119.3              & ---                & --- \\
$T_1$ ($\mu$s)           & $52 \pm 9$         & $22 \pm 1$         & $58 \pm 10$ \\
$T_2^*$ ($\mu$s)         & $31 \pm 4$         & $15.2 \pm 0.8$     & $17 \pm 1$ \\
1Q Fid. (\%)             & $99.73 \pm 0.02$   & $99.70 \pm 0.02$   & $99.84 \pm 0.02$ \\
Readout Fid. (\%)        & $96.8 \pm 0.3$\footnotemark[1]  & $97.15 \pm 0.85$   & $93.65 \pm 2.74$ \\
CZ Fid. (\%)             & N/A                & \multicolumn{2}{c}{$98.46 \pm 0.28$} \\
XY Fid. (\%)             & N/A                & \multicolumn{2}{c}{$98.34 \pm 0.38$} \\
\end{tabular}
\end{ruledtabular}
\footnotetext[1]{Fidelity reported for the $\{\ket{0}, \ket{1}\}$ subspace.}
\end{table}

\section{Experimental Fidelity Data}
\label{app:raw_data}
Table~\ref{tab:fidelities} lists the process fidelities for both implementations across all 40 sampled metrics, including the raw and error-mitigated qutrit results.

\begin{table*}[tbp]
\centering
\caption{Process fidelities for all 40 sampled metrics, showing unmitigated two-qubit results and both raw and error-mitigated qutrit results. Each metric is specified by its minimum eigenvalue $r$ and angular coordinates $(\theta,\phi)$, as defined in Sec.~\ref{subsec:sampling}. An asterisk ($^*$) marks a fidelity strictly above the adversarial threshold $\Delta_{\text{th}}^\eta$. \textbf{Bold} marks the highest measured fidelity for each metric.}
\label{tab:fidelities}
\resizebox{\textwidth}{!}{
\begin{tabular}{@{}cccccccc|cccccccc@{}}
\toprule
\textbf{Index} & \boldmath$r$ & \boldmath$\theta/\pi$ & \boldmath$\phi/\pi$ & \boldmath$\Delta_{\text{th}}^\eta$ & \textbf{Two-qubit} & \textbf{Qutrit (Raw)} & \textbf{Qutrit (Mit.)} & \textbf{Index} & \boldmath$r$ & \boldmath$\theta/\pi$ & \boldmath$\phi/\pi$ & \boldmath$\Delta_{\text{th}}^\eta$ & \textbf{Two-qubit} & \textbf{Qutrit (Raw)} & \textbf{Qutrit (Mit.)}\\ \midrule
1 & 0.905 & 0.75 & -0.50 & 1.000 & $0.967$ & $0.956$ & $\mathbf{0.993}$ & 21 & 0.905 & 0.50 & 1.25 & 1.000 & $0.965$ & $0.934$ & $\mathbf{0.992}$ \\
2 & 0.905 & 0.25 & -0.50 & 1.000 & $0.972$ & $0.967$ & $\mathbf{0.994}$ & 22 & 0.905 & 0.50 & 0.75 & 1.000 & $0.969$ & $0.918$ & $\mathbf{0.984}$ \\
3 & 0.569 & 0.75 & -0.50 & 0.990 & $0.960$ & $0.953$ & $\mathbf{0.988}$ & 23 & 0.905 & 0.50 & 0.25 & 1.000 & $0.960$ & $0.947$ & $\mathbf{0.990}$ \\
4 & 0.569 & 0.25 & -0.50 & 0.990 & $0.966$ & $0.957$ & $\mathbf{0.984}$ & 24 & 0.905 & 0.50 & -0.25 & 1.000 & $0.964$ & $0.959$ & $\mathbf{0.991}$ \\
5 & 0.333 & 0.75 & -0.50 & 0.966 & $0.956$ & $0.901$ & $\mathbf{0.982}^{*}$ & 25 & 0.569 & 0.50 & 1.25 & 0.990 & $0.960$ & $0.924$ & $\mathbf{0.966}$ \\
6 & 0.333 & 0.25 & -0.50 & 0.966 & $0.960$ & $0.945$ & $\mathbf{0.983}^{*}$ & 26 & 0.569 & 0.50 & 0.75 & 0.990 & $0.964$ & $0.927$ & $\mathbf{0.979}$ \\
7 & 0.159 & 0.75 & -0.50 & 0.919 & $0.950^{*}$ & $0.913$ & $\mathbf{0.983}^{*}$ & 27 & 0.569 & 0.50 & 0.25 & 0.990 & $0.957$ & $0.962$ & $\mathbf{0.975}$ \\
8 & 0.159 & 0.25 & -0.50 & 0.919 & $\mathbf{0.959}^{*}$ & $0.908$ & $0.946^{*}$ & 28 & 0.569 & 0.50 & -0.25 & 0.990 & $0.957$ & $0.945$ & $\mathbf{0.987}$ \\
9 & 0.026 & 0.75 & -0.50 & 0.810 & $0.949^{*}$ & $0.926^{*}$ & $\mathbf{0.975}^{*}$ & 29 & 0.333 & 0.50 & 1.25 & 0.966 & $0.951$ & $0.954$ & $\mathbf{0.978}^{*}$ \\
10 & 0.026 & 0.25 & -0.50 & 0.810 & $0.958^{*}$ & $0.872^{*}$ & $\mathbf{0.973}^{*}$ & 30 & 0.333 & 0.50 & 0.75 & 0.966 & $0.956$ & $0.902$ & $\mathbf{0.981}^{*}$ \\
11 & 0.905 & 0.25 & 0.50 & 1.000 & $0.968$ & $0.960$ & $\mathbf{0.987}$ & 31 & 0.333 & 0.50 & 0.25 & 0.966 & $0.952$ & $0.935$ & $\mathbf{0.982}^{*}$ \\
12 & 0.905 & 0.75 & 0.50 & 1.000 & $0.968$ & $0.951$ & $\mathbf{0.981}$ & 32 & 0.333 & 0.50 & -0.25 & 0.966 & $0.953$ & $0.956$ & $\mathbf{0.983}^{*}$ \\
13 & 0.569 & 0.25 & 0.50 & 0.990 & $0.964$ & $0.953$ & $\mathbf{0.975}$ & 33 & 0.159 & 0.50 & 1.25 & 0.919 & $0.943^{*}$ & $0.928^{*}$ & $\mathbf{0.977}^{*}$ \\
14 & 0.569 & 0.75 & 0.50 & 0.990 & $0.963$ & $0.931$ & $\mathbf{0.984}$ & 34 & 0.159 & 0.50 & 0.75 & 0.919 & $0.949^{*}$ & $0.941^{*}$ & $\mathbf{0.980}^{*}$ \\
15 & 0.333 & 0.25 & 0.50 & 0.966 & $0.961$ & $0.952$ & $\mathbf{0.978}^{*}$ & 35 & 0.159 & 0.50 & 0.25 & 0.919 & $0.948^{*}$ & $0.925^{*}$ & $\mathbf{0.969}^{*}$ \\
16 & 0.333 & 0.75 & 0.50 & 0.966 & $0.958$ & $0.918$ & $\mathbf{0.977}^{*}$ & 36 & 0.159 & 0.50 & -0.25 & 0.919 & $0.942^{*}$ & $0.923^{*}$ & $\mathbf{0.978}^{*}$ \\
17 & 0.159 & 0.25 & 0.50 & 0.919 & $\mathbf{0.961}^{*}$ & $0.943^{*}$ & $0.953^{*}$ & 37 & 0.026 & 0.50 & 1.25 & 0.810 & $0.933^{*}$ & $0.894^{*}$ & $\mathbf{0.987}^{*}$ \\
18 & 0.159 & 0.75 & 0.50 & 0.919 & $0.954^{*}$ & $0.944^{*}$ & $\mathbf{0.981}^{*}$ & 38 & 0.026 & 0.50 & 0.75 & 0.810 & $0.935^{*}$ & $0.926^{*}$ & $\mathbf{0.983}^{*}$ \\
19 & 0.026 & 0.25 & 0.50 & 0.810 & $0.956^{*}$ & $0.944^{*}$ & $\mathbf{0.988}^{*}$ & 39 & 0.026 & 0.50 & 0.25 & 0.810 & $0.945^{*}$ & $0.947^{*}$ & $\mathbf{0.979}^{*}$ \\
20 & 0.026 & 0.75 & 0.50 & 0.810 & $0.940^{*}$ & $0.924^{*}$ & $\mathbf{0.973}^{*}$ & 40 & 0.026 & 0.50 & -0.25 & 0.810 & $0.940^{*}$ & $0.930^{*}$ & $\mathbf{0.988}^{*}$ \\ \bottomrule
\end{tabular}
}
\end{table*}

\bibliography{ref}

\end{document}